\documentclass[preprintnumbers, showpacs, floatfix,
preprintnumbers, letterpaper, superscriptaddress, nofootinbib]{revtex4}
\usepackage{amsmath}
\usepackage{amssymb}
\usepackage{graphicx}
\usepackage{color}
\newcommand{\LL}{{\cal L}}
\newcommand{\PP}{{\cal P}}
\newcommand{\be}{\begin{equation}}
\newcommand{\ee}{\end{equation}}
\newcommand{\bea}{\begin{eqnarray}}
\newcommand{\eea}{\end{eqnarray}}
\newcommand{\nn}{\nonumber}

\begin{document}

\title{Model of inflationary magnetogenesis}

\author{Peng Qian}
\email{qianp@itp.ac.cn}
\affiliation{State Key Laboratory of Theoretical Physics, Institute of
Theoretical Physics, Chinese Academy of Sciences, P.O. Box 2735,
Beijing 100190, China}

\author{Zong-Kuan Guo}
\email{guozk@itp.ac.cn}
\affiliation{State Key Laboratory of Theoretical Physics, Institute of
Theoretical Physics, Chinese Academy of Sciences, P.O. Box 2735,
Beijing 100190, China}

\begin{abstract}
We consider the possibility of inflationary magnetogenesis due to dynamical couplings of the electromagnetic fields to gravity.
We find that large primordial magnetic fields can be generated during inflation
without the strong coupling problem, backreaction problem, or curvature perturbation problem,
which seed large-scale magnetic fields with observationally interesting strengths.
\end{abstract}

\maketitle

\section{Introduction}
Magnetic fields are ubiquitous in our Universe and
have been detected in astronomical structures of different sizes, from stars up to galaxy clusters.
In particular, the magnetic field strengths in galaxies and galaxy clusters are typically of the order of $\mu$ gauss.
According to a widely accepted paradigm, the large-scale magnetic fields are produced
by the amplification of initial seeds via the adiabatic compression and turbulent shock flows during structure formation.
To explain the observed magnetic fields in both galaxies and galaxy clusters,
the strengths of the initial seed fields have to be stronger than $10^{-13}$ gauss
on a comoving scale larger than 1 Mpc~\cite{Campanelli:2013mea,Campanelli:2015jfa}.

If the primordial magnetic fields are present before the epoch of photon last scattering,
they will leave significant imprints on the temperature and polarization spectra of the cosmic microwave background (CMB) anisotropies
since their energy-momentum tensor sources all types of cosmological perturbations, {\it i.e.}, scalar, vector, and tensor perturbations.
Therefore, precise measurements of the CMB anisotropies provide constraints
on the primordial magnetic fields~\cite{Durrer:1999bk,Kahniashvili:2000vm,Mack:2001gc,Paoletti:2008ck,Paoletti:2010rx,Paoletti:2012bb,Durrer:2013pga}.
Recently, Planck 2015 data constrained the amplitude of the primordial magnetic fields at a scale of 1 Mpc
to less than a few $10^{-9}$ gauss~\cite{Ade:2015cva}.

So far the origin of the primordial magnetic fields has not yet been completely understood.
Various mechanisms have been proposed to explain them (see Refs.~\cite{Grasso:2000wj,Kandus:2010nw,Yamazaki:2012pg} for some reviews).
The primordial magnetic fields can be produced during cosmological phase transitions,
such as the grand unified theory, electroweak and QCD phase transitions.
The primordial magnetic fields produced during such phase transitions
generally have relevant strengths, but the corresponding correlation lengths are typically too small to explain
large-scale magnetic fields.
It is natural to look for the possibility of generating the primordial magnetic fields on large scales
during inflation in the early Universe.
Since both the Friedmann-Robertson-Walker metric and the electromagnetic fields are conformally invariant,
inflationary magnetogenesis requires the breaking of the conformal invariance
by the coupling of the electromagnetic fields to gravity~\cite{Turner:1987bw,Kunze:2009bs,Kunze:2012rq,BeltranJimenez:2010uh,Jimenez:2010hu},
to the inflaton fields~\cite{Ratra:1991bn,Martin:2007ue,Demozzi:2009fu,Ferreira:2013sqa,Campanelli:2015fna},
to pseudoscalar fields like axions~\cite{Turner:1987bw,Garretson:1992vt,Finelli:2000sh,Campanelli:2008kh},
and to other scalar fields~\cite{Gasperini:1995dh,Bamba:2003av,Bamba:2006ga}.
In principle inflation can provide the dynamical means to amplify electromagnetic quantum fluctuations
that lead to sizable magnetic fields.
Unfortunately, these mechanisms either produce fields that are too small or face serious theoretical problems.
As pointed out in Ref.~\cite{Demozzi:2009fu},
assuming the energy density of the generated electromagnetic fields does not spoil inflationary background dynamics (the backreaction problem),
magnetogenesis models with a weak electromagnetic coupling constant (the strong coupling problem)
cannot explain the origin of the primordial magnetic fields.
Moreover, there exists the curvature perturbation problem
wherein the curvature perturbations produced by the generated electromagnetic fields
are too large to be in conflict with CMB data~\cite{Barnaby:2012tk,Cai:2010uw,Motta:2012rn,Suyama:2012wh,Shiraishi:2013vja,Fujita:2013pgp}.
It was argued in Ref.~\cite{Kanno:2009ei} that inflation still continues even after the backreaction begins to work.
However, it turns out that the creation of primordial magnetic fields is significantly suppressed due to the backreaction.
In sum, these problems make it quite difficult to construct a realistic self-consistent model of inflationary magnetogenesis.

In this paper, we consider the possibility of inflationary magnetogenesis due to dynamical couplings of the electromagnetic
and gravitational fields.
It is known that nonminimal couplings of the electromagnetic fields to gravity may arise from one-loop vacuum polarization in QED in curved space~\cite{Drummond:1979pp}.
Turner and Widrow first applied such couplings to inflationary magnetogenesis and found that the strengths
of the resultant magnetic fields are too weak to explain the observed fields~\cite{Turner:1987bw}.
Here we extend the Turner-Widrow mechanism by assuming that the coupling coefficients are functions of cosmic time.
Such time-dependent couplings are motivated by the dimensional reduction of higher-dimensional theories, {\it e.g.},
string theories, to the four observed spacetime dimensions~\cite{MuellerHoissen:1987ch,Dereli:1990he,Balakin:2008ur}.
With an appropriate choice of the coupling functions, large primordial magnetic fields can be generated during inflation
without the strong coupling problem, backreaction problem, or curvature perturbation problem,
to seed the large-scale magnetic fields with observationally interesting strengths.

\section{Model}\label{sec:model}
We consider the action for the electromagnetic field $A_\mu$,
\begin{equation}
S=\int d^4x \sqrt{-g}\left[\frac12 R+\LL_{\rm inf}-\frac{1}{4}I_1F^2+I_2RF^2+I_3R_{\mu}{}^{\nu}F^{\mu\rho}F_{\nu\rho}+I_4R_{\mu\nu\rho\sigma}F^{\mu\nu}F^{\rho\sigma}\right],
\label{eq:action}
\end{equation}
where $\LL_{\rm inf}$ is the inflation Lagrangian density and $I_i (i=1,...,4)$ are functions of the inflaton, dilaton, or some other scalar field $\phi$.
In the case of $I_1=1$ and $I_2=I_3=I_4=0$, the standard electromagnetic action is recovered,
in which the electromagnetic field fluctuations can not be amplified during inflation
because the action is conformally invariant in the Friedmann-Robertson-Walker Universe.
Inflationary magnetogenesis requires the breaking of the conformal invariance of the electromagnetic action.
Here the couplings $I_i$ are considered for this purpose.
Unless explicitly stated, we work with natural units in which $c=\hbar=1$.
From the action~\eqref{eq:action}, the equation of motion for the electromagnetic field is given by
\begin{equation}
\nabla_{\mu}\left[-I_1F^{\mu \nu }+4I_2{RF}^{\mu \nu }+2I_3(R^{\rho \mu }F_{\rho }{}^{\nu }-R^{\rho \nu }F_{\rho }{}^{\mu })+4I_4R^{\mu \nu \rho \sigma }F_{\rho \sigma }\right]=0.
\end{equation}
In principle, for a given form of the expressions for the couplings $I_i$ we can consider the possibility of inflationary magnetogenesis.
As an illustration, we assume that $I_i$ have the same dependence on $\phi$, which is a dilatonic extension of the Turner-Widrow model~\cite{Turner:1987bw}, so that
\begin{eqnarray}
I_1=c_1f^2(\phi),\quad I_2=c_2f^2(\phi),\quad I_3=c_3f^2(\phi),\quad I_4=c_4f^2(\phi),
\end{eqnarray}
where the coefficients $c_i$ are constants. We consider a spatially flat Friedmann-Robertson-Walker metric
\bea
ds^2 = a^2(\eta)(-d\eta^2+d\mathbf{x}^2),
\eea
where $\eta$ is the conformal time and $a$ is the scale factor.
We choose the Coulomb gauge, $A_0=0$ and $\partial_{i}A^{i}=0$, and
expand $A_i$ as
\begin{eqnarray}
A_i(\eta, \mathbf{x}) = \sum_{\sigma=1,2}\int\frac{d^3k}{(2\pi)^{3/2}}
 \left(\epsilon_{i,\sigma}(\mathbf{k}) a_{\mathbf{k},\sigma} A_{k}(\eta) e^{i\mathbf{k}\cdot\mathbf{x}}+{\rm H.c.}\right),
\end{eqnarray}
where $\epsilon_{i,\sigma}(\mathbf{k})$ are the two orthonormal transverse polarization vectors,
and $a_{\mathbf{k},\sigma}$ and $a_{\mathbf{k},\sigma}^{\dagger}$ are the annihilation and creation operators
which satisfy the commutation relations
$[a_{\mathbf{k},\sigma},a_{\mathbf{k}',\sigma'}^{\dagger}]=\delta_{\sigma\sigma'}\delta(\mathbf{k}-\mathbf{k}')$
and $[a_{\mathbf{k},\sigma},a_{\mathbf{k}',\sigma'}]=[a_{\mathbf{k},\sigma}^{\dagger},a_{\mathbf{k}',\sigma'}^{\dagger}]=0$.
In an exact de Sitter background, the Fourier modes $A_k$ satisfy the equation
\begin{eqnarray}
\chi^2  \left(A_k''+\frac{2f'}{f} A_k'+k^2A_k\right)=0,
\label{eq:me}
\end{eqnarray}
with the normalization condition
\bea
A_{k}(\eta)A_{k}'^{*}(\eta)-A_{k}^{*}(\eta)A_{k}'(\eta)=\frac{i}{\chi^2 f^2}\,,
\eea
where a prime denotes a derivative with respect to $\eta$.
Here, we have defined $\chi^2=c_1-4 (12 c_2+3 c_3+2 c_4)H_I^2$,
where $H_I$ is the value of the Hubble parameter during inflation.
To have positive kinetic terms we must require
\bea
c_1-4 (12 c_2+3 c_3+2 c_4)H_I^2>0,
\label{eq:ghostfree}
\eea
so that the theory is ghost free~\cite{Jimenez:2013qsa}.
In terms of a new variable $v_k=\chi fA_k$, Eq.~\eqref{eq:me} becomes
\begin{eqnarray}
\chi \left[v_k''+\left(k^2-\frac{f''}{f}\right)v_k\right]=0.
\label{eq:newme}
\end{eqnarray}

\subsection{Strong coupling problem}
It is assumed that classical electromagnetism is restored at the end of inflation, and therefore $I_1\to 1$.
As pointed out in Ref.~\cite{Demozzi:2009fu}, the theory is not trustworthy in the strong coupling regime of
$I_1 \ll 1$ during inflation
because $1/I_1$ is an effective coupling constant between the electromagnetic field and a charged fermion.
This is called the strong coupling problem.
A natural solution to such a problem is to require $I_1>1$ during inflation.

We assume that $f$ is a power-law function of the scale factor as
\begin{eqnarray}
f=f_e\left(\frac{a}{a_e}\right)^n,
\label{eq:ff}
\end{eqnarray}
where $a_e$ is the value of the scale factor at the end of inflation, $f_e$ is the value of $f$ at the end of inflation, and $c_1 f^2_e\approx{\cal{O}}(1)$. The form~\eqref{eq:ff} is valid during inflation.
After inflation the standard electromagnetism is recovered.
In what follows we shall consider the weak-coupling case of $n<0$.
If $\chi \neq 0$,
for short waves with $k^2\gg f''/f$ the solution of Eq.~\eqref{eq:newme} is
\begin{eqnarray}
v_k(\eta)=\frac{1}{\sqrt{2k}}e^{-ik\eta},
\label{eq:inis}
\end{eqnarray}
and for long waves with $k^2\ll f''/f$ we have the general solution
\begin{eqnarray}
v_k(\eta)=C_1 a^n + C_2 a^{-n-1},
\end{eqnarray}
where $C_1$ and $C_2$ are integration constants which can be fixed by matching this solution to Eq.~\eqref{eq:inis}.
If $-1/2<n<0$, the first mode dominates over the second one.
Since $v_k\propto a^n$, $A_k$ is constant during inflation.
In this case, the power spectrum of the primordial magnetic fields decreases towards large scales,
which corresponds to a strong blue spectrum.
Therefore, it is hard to produce a large amplitude of the magnetic fields on Mpc scales~\cite{Demozzi:2009fu}.
In the case of $n<-1/2$, the second mode dominates, {\it i.e.}, $v_k\propto a^{-n-1}$.
Using the vacuum initial condition~\eqref{eq:inis}, we have
\begin{equation}
A_{k}(\eta)=\frac{1}{\chi f_e\sqrt{2k}}\left(\frac{aH_I}{k}\right)^{-n-1} \left(\frac{a}{a_e}\right)^{-n}.
\end{equation}
Notice that $A_k$ grows rapidly during inflation, which leads to the backreaction problem.
In the next subsection we shall discuss it.
The power spectrum of the primordial magnetic fields is
\begin{equation}
\PP_B(k,\eta)=\sum_{\sigma=1,2}\frac{k^5|A_k(\eta)|^2}{2\pi^2a^4}
=\frac{H_I^4}{2\pi^2\chi^2 f_e^2}\left(\frac{k}{aH_I}\right)^{2n+6}\left(\frac{a}{a_e}\right)^{-2n},
\label{eq:ps}
\end{equation}
with the spectral index $n_B=2n+6$. The spectrum is scale invariant for $n=-3$,
in which the amplitude at the end of inflation is determined by $H_I$.
Using the entropy conservation we have $a_0/a_e\sim \sqrt{H_I/m_{pl}}/(k_B T_0/m_{pl})\sim 10^{32}\sqrt{H_I/m_{pl}}$
for instantaneous reheating of inflation,
where $a_0$ and $T_0$ are the scale factor and CMB temperature today, $k_B$ is the Boltzmann constant and $m_{pl}$ is the Planck mass.
From Eq.~\eqref{eq:ps} the strength of the magnetic fields $B_\lambda$ at a scale of $\lambda_0=a_0/k$ in Mpc is
\be
B_\lambda \sim 10^{58} \left( \frac{H_I}{m_{pl}}\right)^2 \left(\frac{a_0}{a_e}\right)^{n+1}\left(\frac{\lambda_0}{1 {\rm Mpc}}\frac{10^{58}}{(H_I/m_{pl})^{-1}}\right)^{-n-3} {\rm gauss}.
\ee
For $n=-3$ and $H_I\sim 10^{-6} m_{pl}$, the amplitude of the magnetic fields is about $10^{-12}$ gauss on all scales today.
Compared with the scale-invariant spectrum, for a blue spectrum with $n>-3$ the amplitude on Mpc scales is suppressed
while for a red spectrum with $n<-3$ it is enhanced.
Here, we have assumed that the Universe is always a good conductor and ignored dissipative effects.
It is known that a finite electric conductivity dramatically damps the fields.
Compared with the expansion of the Universe, the dissipative effects are rather smaller during most periods of evolution of the Universe.
More details have been discussed in Ref.~\cite{Campanelli:2015jfa}.

\subsection{Backreaction problem \label{subsection:back}}
In successful inflationary magnetogenesis scenarios, the contribution of the electromagnetic fields
to the background energy density is negligible
so that the background dynamics is controlled by the inflaton.
Here we shall show that the backreaction problem can be avoided in some parameter space.
The energy-momenta tensor of the electromagnetic fields is given by
\begin{eqnarray}
T^{\mu }{}_{\nu } &=& I_1\left(F^{\mu }{}_{\rho }F_{\nu }{}^{\rho }-\frac14 g^{\mu }{}_{\nu }F^2\right)
  - I_2\left(4RF^{\mu \rho }F_{\nu \rho }+2R^{\mu }{}_{\nu }F^2-g^{\mu }{}_{\nu }RF^2\right) \nn \\
  && - I_3\left(4R^{\mu }{}_{\sigma }F_{\nu \rho }F^{\sigma \rho }
  + 2R^{\rho \sigma }F_{\rho }{}^{\mu }F_{\sigma \nu }
  - g^{\mu }{}_{\nu }R^{\rho \sigma }F_{\rho \lambda}F_{\sigma }{}^\lambda\right) \nn \\
  && - I_4\left(8R^{\mu \lambda }{}_{\rho \sigma }F_{\nu \lambda }F^{\rho \sigma }
  - g^{\mu }{}_{\nu }R_{\lambda \gamma \rho \sigma }F^{\lambda \gamma }F^{\rho \sigma }\right).
\end{eqnarray}
The energy density of the electromagnetic fields then is the vacuum expectation value of the $T^0{}_0$ component,
\begin{eqnarray}
  \rho_{EM} =-\langle T^0{}_0\rangle
  = f^2\int \frac{dk}{k}(Q_1\PP_E+Q_2\PP_B),
\label{eq:energy}
\end{eqnarray}
where
\begin{eqnarray}
&& Q_1=c_1 - 24(3c_2+c_3+c_4)H_I^2, \\
&& Q_2=c_1 - 4(6c_2+3c_3+2c_4)H_I^2,
\end{eqnarray}
and the power spectrum of the electric fields is defined as
\bea
\PP_E =\sum_{\sigma=1,2}\frac{k^3|A'_k(\eta)|^2}{4\pi^2 a^4}
 = \frac{(2n+1)^2H_I^4}{2\pi^2\chi^2 f_e^2}\left(\frac{k}{aH_I}\right)^{2n+4}\left(\frac{a}{a_e}\right)^{-2n}.
\eea
For the trace of the $T^{i}{}_{j}$ components of the energy-momentum tensor, we have
\begin{eqnarray}
\langle T^{i}{}_{i}\rangle
 = f^{2} \int \frac{dk}{k}( Q_3\PP_E+Q_4\PP_B),
\label{eq:trace}
\end{eqnarray}
where
\begin{eqnarray}
&& Q_3=c_1 + 8(3c_2+c_4)H_I^2,\\
&& Q_4=c_1 - 4(30c_2+9c_3+10c_4)H_I^2.
\end{eqnarray}
At the end of inflation $\PP_B\propto (k/a_e H_I)^{2n+6}$, while $\PP_E\propto(k/a_e H_I)^{2n+4}$ with $k<a_e H_I$ in the weak-coupling regime with $n<-1/2$.
Hence, the main contribution to the energy density~\eqref{eq:energy} and the pressure density~\eqref{eq:trace}
comes from the power spectrum of the electric fields.
As found in Ref.~\cite{Demozzi:2009fu} for the special case of $c_2=c_3=c_4=0$,
in order to produce the magnetic fields with sufficient strengths during inflation,
the backreaction of the electric fields would spoil inflationary dynamics.
If $\rho_{EM}<H_I^2$ is required, the produced fields are too weak to explain the origin of the primordial magnetic fields.
However, in our model thanks to the time-dependent couplings of the electromagnetic fields to gravity,
we can neglect the contribution of the electric fields to the background energy density
if $Q_1=Q_3=0$, i.e.,
\begin{eqnarray}
\label{eq:q1}
&& \frac{c_3}{c_1}=\frac{1}{6H_I^2}\,,\\
&& 3\frac{c_2}{c_1}+\frac{c_4}{c_1}=-\frac{1}{8H_I^2}\,.
\label{eq:q3}
\end{eqnarray}
Therefore, the backreaction problem can be avoided in some parameter space.

In the standard kinetically coupled scenario for magnetogenesis,
the electromagnetic energy density is the integration of the power spectra with respect to $k$.
Therefore, it is hard to balance the magnetic strengths and electric energy density.
In the presence of the couplings of the electromagnetic fields to gravity,
the relation between the power spectra and energy density depends on the coupling coefficients.
Canceling the contribution of the electric field to the background energy density
implies that some of the coupling terms in Eq.~\eqref{eq:action} provide equivalently
negative energy densities, but the total energy of the electromagnetic fields is still positive.
As a phenomenological model for magnetogenesis,
we introduce the fine-tuning of the coupling coefficients.

\subsection{Curvature perturbation problem}
It is known that in single-field slow-roll inflationary scenarios quantum fluctuations are frozen
when the wavelengths of the fluctuations are well outside the Hubble horizon.
In the presence of the electromagnetic fields produced during inflation,
the curvature perturbation $\zeta$ still evolves on super-Hubble scales.
In this case we have to calculate the curvature perturbation at the end of inflation.
The evolution of $\zeta_{EM}$ on super-Hubble scales is~\cite{Wands:2000dp,Lyth:2004gb,Barnaby:2012tk,Campanelli:2015jfa}
\begin{eqnarray}
\dot{\zeta}_{EM}=-3(1+w_{EM})\frac{H_I}{2\epsilon\rho_I}\rho_{EM},
\end{eqnarray}
where a dot denotes a derivative with respect to the cosmic time,
$w_{EM}$ is the electromagnetic equation of state,
and $\epsilon$ is the slow-roll parameter~\cite{Stewart:1993bc,Schwarz:2001vv,Leach:2002ar,Finelli:2009bs}.
Assuming $H_I$ and $\epsilon$ are constant during inflation,
we obtain the curvature perturbation induced by the electromagnetic fields
\begin{eqnarray}
\zeta_{EM} = -3(1+w_{EM})\frac{H_I}{2\epsilon \rho_I}\int_{t_{i}}^{t_e} dt \rho_{EM}(t),
\label{eq:perturbations}
\end{eqnarray}
where $t_i$ is an initial time when $\zeta_{EM}=0$ and $t_e$ is the time at the end of inflation.

In Sec.~\ref{subsection:back}, it was concluded that $Q_1 = 0$ cancels the contribution of the
electric fields to the background energy density.
Now we estimate the contribution of the magnetic fields to the curvature
perturbation.
In the case of $n=-3$, by substituting Eq.~\eqref{eq:ps} into Eq.~\eqref{eq:energy}
we have $\rho_{EM} = N Q_2 H_I^4/(2 \pi^2 \chi^2)$, where $N$ is the number of $e$-folds of inflation.
The curvature perturbation induced by the magnetic fields is approximated by $\zeta_{EM} \sim N^2 Q_2 H_I^2/(m_{pl}^2 \epsilon \chi^2)$,
which leads to an amplitude of the power spectrum $\PP_\zeta^{EM} \sim N^4 Q_2^2 H_I^4/(m_{pl}^4 \epsilon^2 \chi^4)$.
It is known that the power spectrum of curvature perturbations generated by the inflaton is given by
$\PP_\zeta^I \sim H_I^2/(m_{pl}^2\epsilon)$.
We show that the dominant component of the power spectrum of curvature perturbations
is generated by the inflaton, {\it i.e.}, $\PP_\zeta^I > \PP_\zeta^{EM}$.
For $H_I\sim 10^{-6}m_{pl}$ and $N = 60$, the amplitude of the power spectrum with $\epsilon \sim 10^{-2}$
is consistent with Planck 2015 data~\cite{Ade:2015lrj,Ade:2015xua}.
In this case, we have $\PP_\zeta^{EM}\sim 10^{-13}Q_2^2/\chi^4$, which is smaller than $\PP_\zeta^I$ if $Q_2/\chi^2 \lesssim 10^2$.
Therefore, the contribution of the electromagnetic fields to the curvature perturbation
can safely be neglected if $Q_2/\chi^2 \lesssim 10^2$.

\section{Conclusion}\label{sec:conclusion}
In this paper, we have considered a model of inflationary magnetogenesis based on the Turner-Widrow mechanism,
in which the conformal invariance of the electromagnetic fields is broken by the nonminimal couplings of the electromagnetic fields to gravity.
A scale-invariant power spectrum of primordial magnetic fields can be generated during inflation in the weak-coupling regime,
which can seed the large-scale magnetic fields with observationally interesting strengths on the Mpc scale.
Since the relation between the energy density of the electromagnetic fields and their power spectra depends on the coupling coefficients,
with appropriate choices of the coefficients, {\it i.e.}, Eq.~\eqref{eq:ghostfree}, \eqref{eq:q1}, and \eqref{eq:q3},
the strong coupling problem and curvature perturbation problem can be avoided in our model.

\begin{acknowledgements}
This work is supported by the National Natural Science Foundation of China Grants No.11575272 and No.11335012.
\end{acknowledgements}


\end{document}